\documentclass[preprint,aps,prb,epsf]{revtex4}   
\include{amsmath}
\include{epsfx}
\usepackage{graphicx}
\begin{document}
\title{{\bf Stretching of Proteins in a Uniform Flow}}
\author{{\bf P. Szymczak $^1$ and Marek Cieplak$^2$}}

\address{
$^1$Institute of Theoretical Physics, Warsaw University,
ul. Ho\.za 69, 00-681 Warsaw, Poland\\
$^2$Institute of Physics, Polish Academy of Sciences,
Al. Lotnik\'ow 32/46, 02-668 Warsaw, Poland}

\vskip 40pt
\noindent {\bf
Keywords: protein stretching, protein folding, manipulation of
proteins, Go model,
molecular dynamics, ubiquitin, integrin}

\noindent {PACS numbers: 82.37.Rs, 87.14.Ee, 87.15.-v}

\vspace*{1cm}

\begin{abstract}
{Stretching of a protein by a fluid flow is compared to that in a
force-clamp apparatus. The comparison is made
within a simple topology-based dynamical model of a protein in
which the effects of the flow are implemented using Langevin dynamics.
We demonstrate that unfolding induced by a uniform flow shows a richer
behavior than that in the force clamp. The dynamics of unfolding
is found to depend strongly on the selection of the amino acid,
usually one of the termini, which is anchored. These features offer
potentially wider diagnostic tools to investigate structure of proteins
compared to experiments based on the atomic force microscopy.}
\end{abstract}

\maketitle

\section{Introduction}

The deformation of polymers in a flow has been a subject of active research for
at least seventy years (see e.g. \cite{kuhn,kramers,gennes}).
A recent renewed interest in this topic \cite{smith,perkins1,perkins2,rzehak,cheon}
 arose due to
development of precise experimental techniques allowing for studies of
conformations
at a single-molecule level. In particular, biological macromolecules
such as DNA have been intensely studied in this way
\cite{larson,abramchuk,shaqfeq,blossey}.
Stretching by a flow is also at the heart of the manipulation technique known as
molecular combing used in genomic studies
\cite{Strick,Bensimon,BensBens,Austin1,Austin}
and in nano-electronics \cite{Inganas}.\\

In this paper, we concentrate on analysis of protein unfolding in uniform flow
and compare it with unfolding in a force clamp \cite{clampober}, i.e.
under the condition of a constant force applied to one of the termini.
Theoretical studies on protein stretching in a flow are scarce
\cite{lemak1,lemak2} and
limited to the minimalist $\beta$ - barrel model.
Here, we present a theoretical method to study flow induced
deformations of, in principle, any protein and we illustrate it by
considering ubiquitin and integrin. These two proteins were chosen
because of their contrasting dynamical behavior, as established through
simulations \cite{cs1}, when unfolding in a force clamp: 
ubiquitin unfolds as a rule in a single kinetic step whereas integrin
-- in multiple steps, i.e. with several intermediates. Additionally,
we also consider synthetic $\alpha$-helices, which are homopolymers,
and show a very different behaviour than that seen in complex proteins.\\

The model we propose here is coarse grained and it involves no explicit solvent.
It is an extension of the Go-like modelling \cite{Goabe} that we have
employed in the past
to study folding \cite{Hoang,biophysical} and stretching at constant velocity
\cite{thermtit,Pastore,ubiq}. In short, a protein is represented by
a chain of C$^{\alpha}$ atoms that are tethered by harmonic potentials
with minima at 3.8 {
\AA}.
The effective self-interactions between the atoms are either purely
repulsive or are
minimum-endowed-contacts of the Lennard-Jones type. The parameters of
these potentials
reflect existence of the surrounding solvent in the sense that their minima
correspond to the experimentally determined distances between the
C$^{\alpha}$ atoms of a
protein in water.\\

The dynamics of the protein is governed by the Langevin equation
\begin{equation}
m \ddot{{\bf r_i}} = -\gamma (\dot{{\bf r}}_i-{\bf u}({\bf r}_i)) +
F^c_i + \Gamma \;\;.
\label{lang}
\end{equation}

Here, ${\bf r_i}$ is the position of $i$'th aminoacid, $F^c_i$ is the
net force on it
due to contact potentials, $\gamma $ is the friction coefficient, and
$\Gamma$ is a white
noise term
with the dispersion of $\sqrt{2\gamma k_B T}$, where $k_B$ is the
Boltzmann constant. Finally,  ${\bf u}({\bf r}_i)$ denotes the solvent
flow field.\\

In principle the model allows for more general description in which
both $\gamma$ and $m$
are amino acid dependent and, in particular, friction is reduced
for hydrophobic residues since hydrophobicity leads
to a slip \cite{knudsenprl,lotus}. In this paper, we stay with the spirit
of the traditional Go-like modelling in which all features of the
protein are assumed to be contained in the native geometry and consider
uniform masses and friction coefficients. Also, we neglect the effects of hydrodynamic
interactions,
which corresponds
to the free- draining limit. This is a serious approximation, since the
hydrodynamic forces between the particles contain long-range terms  decaying as
$R^{-1}$ with the interparticle distance. However, since the number of residues
in the considered proteins is relatively large (100- 200), the inclusion of
hydrodynamic interactions (HI) into the Langevin dynamics scheme would add
considerably to the numerical complexity of the problem, rendering an accurate
calculation of mean unfolding times unfeasible, particularly in the small-force
regime where those times are exceedingly long.\\

In fact, the hydrodynamic effects are very rarely taken into account in the
numerical simulations of protein folding and unfolding: in the all-atom MD
simulations sometimes an explicit solvent is used; but this restricts severely
the largest feasible trajectory length. The coarse-grained models, in
principle, would be a best starting point for the analysis of the impact of HI
on the protein dynamics. However, to the best knowledge of the authors, no
detailed studies of the impact of HI on protein folding and unfolding were
performed, even though it was argued \cite{Tanaka} that such an effect is
expected to be non-negligible.\\

Keeping the above in mind, we nevertheless believe that the free-draining case
may still provide useful insights on protein dynamics in a flow. This is
partially confirmed by the results of the analysis of  DNA streching in a
uniform flow \cite{perkins2}, where it has been observed that the change in the
extension vs flow dependence is very modest when no hydrodynamic interactions
are included in the model (see also the respective discussion in
\cite{shaqfeq}). A similar conclusion is drawn by Hsieh et al. \cite{hsieh} who
compare the results of extensive simulations of the bead-spring DNA models in
extensional flow to the experimental data by Perkins et al. \cite{perkins1}.
They note that deformation-dependent HI has very little effect on the
extensional flow properties of DNA molecules, whereas the rates of unraveling
of single long molecule of DNA observed optically in an extensional flow can be
even quantitatively predicted by bead–spring models that neglect HI.\\

Naturally, due to their highly heterogeneous structure, proteins are much more
complex than a DNA chain, thus one cannot expect those results to apply
directly to the protein streching in a flow. Still, it seems that a free-
draining case may provide a good starting point for understanding, at least on
the qualitative level, the properties of the protein in the flow.\\

With the use of this simplified model, we demonstrate that flow may stretch
proteins to partially unravelled stationary conformations that depend on the
flow rate and on the selection of the terminus which is anchored. This is in
contrast to stretching in a force clamp, in which the set of intermediate
states stays the same whether we fix the C terminus and pull on the N one or do
it the other way around. This difference is caused by the fact that a flow
generates a non-uniform tension in the polymer.
A simple explanation of this phenomenon is
presented in Figure 1 for the case of a linearly positioned chain of $N$ beads connected by
bonds
(in the absence of hydrodynamic interactions):
the bond which is most distant from the anchor is pushed by $1/N$ of the
force that is experienced by the first bond since the latter accumulates all
individual pushes. It follows that, as the flow velocity is increased,
the contacts near the anchored end of the protein are broken first and
the protein
unwinds segment by segment starting from the fixed end. Similar phenomena are
observed in the experiments and simulations of polymers subject to a uniform
flow (see \cite{rzehak} and references therein) and the corresponding
shape of the partially
unwound polymer was called
``stem and a flower''\cite{bro1,bro2} or ''ball and a string"
\cite{Haupt,Maurice,homop}.
Thus exploration of stationary conformations corresponding to various flow
rates should offer a more telling diagnostic of elastic properties of
the protein than the
force clamp.\\

We present the model in Section 2 and consider the case of the helix in
Section 3. Stretching in a constant flow is analyzed in Section 4 for
various ways of
choosing
the anchoring point and for tandem arrangement of proteins.
In Section 5, we discuss an example of a non-uniform flow: elongational one.
Finally, in Section 6, we consider refolding after stopping the flow.\\

\section{The model}

The effective interactions between the C$^{\alpha}$ amino acids are
split into two classes: native and non-native. The distinction
is done by checking for native overlaps of all atoms in aminoacids when
represented by enlarged van der Waals spheres as proposed in
reference \cite{Tsai}. The amino acids, $i$ and $j$ that do overlap in
this sense
are endowed with the effective Lennard-Jones potential
$V_{ij} = 4\epsilon \left[ \left( \frac{\sigma_{ij}}{r_{ij}}
\right)^{12}-\left(\frac{\sigma_{ij}}{r_{ij}}\right)^6\right]$.
The length parameters $\sigma _{ij}$ are chosen so that the potential
minima correspond, pair-by-pair, to the experimentally established
native distances between
the C$^{\alpha}$ atoms in amino acids in the pair. The repulsive
interactions are described
by the $r_{ij}^{-12}$ part of the Lennard-Jones potential combined
with a constant shift term
that makes the potential vanish smoothly at $\sigma =5$ {\AA}.
It should be noted that the specificity of a protein is contained in
the length parameters
$\sigma _{ij}$. The energy parameter, $\epsilon$, is taken to be
uniform and its effective
value
for titin and ubiquitin appears to be of order 900 K so the reduced temperature,
$\tilde{T}=k_BT/\epsilon \;$ of 0.3
($k_B$ is the Boltzmann constant and $T$ is temperature)
should be close to the room temperature
value \cite{Pastore,ubiq}. All of the simulations
reported here were performed at this temperature.
In our stretching simulations, the anchored terminus of the protein is
attached to a harmonic spring of elastic
constant $k$=0.06 $\epsilon /${\AA}$^2$.\\

As explained in the Introduction,
thermostating and mimicking some effects of the solvent
are provided by the Langevin dynamics, Eq. 1. 
The friction coefficient $\gamma$ is taken to be equal to $2m/\tau$
where $\tau =\sqrt{m \sigma^2 / \epsilon} \approx 3 ps$
is the characteristic time scale of oscillations in the Lennard-Jones well.
The selected value of $\gamma$ corresponds to a situation in which the
inertial effects are small \cite{Hoang} but the damping action is not
yet as strong as in water.\\

The equations of motion are solved by a fifth order predictor-corrector scheme.
In the course of stretching, the native contacts are being ruptured. A
contact between
amino acids $i$ and $j$ is said to be ruptured if the corresponding
distance $r_{ij}$ becomes
larger
than $1.5 \sigma _{ij}$ (close to the inflection point of the
Lennard-Jones potential) for
the last time.
When studying folding, we consider establishing contacts
starting from an unfolded state. A contact is said to be established
when the corresponding value of $r_{ij}$ crosses the threshold value
for the first
time. Folding is considered to be achieved when {\em all} contacts are
established simultaneously.\\

When simulating the force clamp, the force $F$ is applied to the spring that
pulls one of the termini (the choice of the terminus is irrelevant in
this case).
In the case of the flow, we discuss the results in terms of the net
hydrodynamic force, ${\bf F}=\gamma \sum_{i=1}^{N} {\bf u}({\bf r}_i)$, that is
experienced at the anchor point. For uniform flows, $F=N\gamma u$. The
dimensionless force, $F\sigma/\epsilon$, will be denoted by $\tilde{F}$.
The conformations will be characterized by the end-to-end
distance $L$.\\

The relative strength of convective and diffusive effects in the dynamics of a
protein is given by the Peclet number $$ Pe = \frac{U R_g}{D} $$ where $U$ is
the characteristic flow magnitude, $R_g$ - radius of gyration and $D$ - the
diffusion coefficient of the protein. Numerically, one may estimate $D$ by the
analysis of the mean square displacement of the protein as a function of
time. For example, for ubiquitin, the calculations give
$D \approx 0.2  {\sigma^2 / \tau}$, whereas $R_g=2.3 \sigma$. The flow rates
used in the simulations lie in the range $U=0.02 - 0.07 \sigma / \tau$ ,
what gives $\mbox{Pe} \approx 0.2 - 0.7$.\\

Since the Peclet number is dimensionless, it can be used to relate the
simulation to the experimental setup. Namely, as reported in \cite{diff}, the
diffusion coefficient for ubiquitin is $D \approx 1.7 \cdot 10^{-6}$ cm$^2$/s, whereas
$R_g=1.15 \cdot 10^{-7} cm$. Thus the above mentioned Peclet number range
corresponds to the flow rates of $4-13$ cm/s. Such speeds are about three
orders of magnitude faster than those needed to unravel DNA molecules. This is
because proteins contain larger clusters of bonds that need to be ruptured
simulateneously and are also smaller in size.
The above comparison may also be used to relate the numerical time scale $\tau$ to the
experimental time scales.
Namely, from the fact that $U=0.02 \sigma / \tau$ corresponds to
4 cm/s and $\sigma = 5\; \AA$ one concludes that  $\tau$ corresponds to approximately
0.25 ns of the `real' time. This time scale is of the same order of magnitude
as the one that Veitshans et al. \cite{veit} arrived at (3 ns) by using an entirely
different argument.\\

\section{Flow-induced stretching of homopolymers}

A synthetic helix provides an example of a homopolymer since none of its
parts, except at the termini, is distinguished.
The dependence of the end-to-end distance in the stretched helix on the total
stretching force is shown in Figure 2. The fractional extension is seen to be a
smooth function of the applied force without any stationary or quasistationary
stages.
The steady-state conformations corresponding to different values of $F$ clearly
show the ``stem and a flower'' phenomenon -- the helix unwinds from
the fixed end
and the unwound length depends on the net hydrodynamic force, i.e. on
the flow rate.
It is also seen that the dependence of the fractional change in $L$ on $F$
does not change with the total number of residues in the helix.
This finding is consistent with a similar and well-established result
\cite{rzehak,bro1,bro2} pertaining to homopolymers in the free-draining limit.\\
The other part of Figure 2 shows the dependence of the mean unfolding
time of the helix
on the net hydrodynamic force.  For the purpose of making this figure,
we consider
the helix to be unfolded when its total length exceeds 90\% of the
maximum extension length
of $(N-1) \times 3.8 \AA$. It is seen that in the small-force regime,the
unfolding time exponentially decreases
as a function of the force. For larger forces the dependence
of the unfolding time on the force becomes much weaker. An analogous
phenomenon is observed in the
simulations of stretching of proteins in the force clamp \cite{cs1}.

\section{ Flow-induced stretching of proteins}

Figures 3 and 4 show the end-to-end distance versus time for  integrin
unfolding in a uniform flow. In
Figure 3, terminus C is anchored, whereas in Figure 4 it is terminus N
that is anchored. Several
trajectories corresponding to different values of the total
hydrodynamic force are shown.  Since the
tension is strongest near the anchoring point, the protein unfolds
from the fixed end towards the free
one. In contrast to homopolymer, here the unfolding pathway traverses
through a number of intermediate
states, corresponding to the unzipping of subsequent structures from
the bulk of the protein. We observe
that if the flow rate is sufficiently low, the protein may remain
trapped in one of these states for the
duration of the simulation. In contrast to the simple helix, complex
proteins are cross-linked and
inhomogeneous and yield to inhomogeneous tension in a way which is
specific to the stretching protocol. Thus the
steady-state conformation in which the protein is found after a long
time depends not only on the value
of the force but also on the choice of the terminus. In particular, as
it is seen in the Figures, the set
of intermediates is much richer for the case of fixed C terminus than
vice versa. Also, in the former
case the full unwinding of integrin chain requires a smaller net
force. This suggests that the strongest
bonds in the native structure of integrin are located nearer to the C
terminus.\\

 The differences between
unfolding with different termini fixed are further highlighted by
analysis of the so called unfolding scenarios
\cite{Hoang}, in which one plots an average time when a given contact
is broken against the contact
order, i.e. against the sequential distance, $|j - i|$, between the
amino acids that form a native
contact. Figure 5 compares the unfolding scenarios for different
anchorings of the integrin chain. Again,
it is seen that anchoring at the C terminus gives rise to a much
richer unfolding dynamics, including
several intermediates, than the anchoring at the other terminus.
Existence of these differences may offer
an interesting way of the experimental probing the structure of a
protein by analysis of unfolding
trajectories with different anchoring points.\\

For comparison, Figure 6 shows the unfolding trajectories for the
integrin pulled by a force applied at
the terminus only, as in the force-clamp apparatus. All the
intermediates present here are also seen in
uniform flow experiments, particularly those with C terminus fixed
({\it cf.} Figure 3), but vice versa
is not true. Thus uniform flow unfolding appears to be richer in intermediate
conformations than simple pulling.\\

It is instructive to perform a similar analysis for another protein,
ubiquitin. Ubiquitin
behaves very differently from integrin when stretched in a force-clamp
\cite{cs1}, since it
unfolds usually in a single kinetic step whereas integrin unfolding
involves several
intermediates as it is illustrated in Figure 5. However, as it is seen
in Figures 7 and 8, in a
uniform flow ubiquitin does display several intermediate steps in
unfolding, which confirms
that the unfolding in a flow shows a richer behavior than stretching
in a force clamp.
Additionally, one again observes a difference in behavior between $C$
and $N$ anchoring. This
time, it is $N$ anchoring which shows a larger number of intermediates
and requires a smaller
force for the full unfolding. Thus, in the case of ubiquitin, the
strongest bonds are located
in the neighborhood of $N$ terminus.\\

Finally, Figure 9 summarizes results on statistically averaged (over
50 unfolding
trajectories) flow-induced and force-clamp-induced processes of
unfolding in ubiquitin (top
panel) and integrin (bottom panel). The figure shows the dependence of
the logarithm of the
median unfolding time on the total force. Just as in the case of
helix, we consider
a protein to be unfolded when its total length exceeds 90\% of the
maximum extension length
of $(N-1) \times 3.8 \AA$. Again, for the uniform flow, one should note
the lack of symmetry
between the anchoring by the N terminus and by the C terminus.
One should also observe that determination of
which choice offers more resistance to unravelling is protein dependent.
However, force-clamp
stretching generally requires a smaller force since in the
flow-induced case the segments
which are near the free end are exposed to relatively small
unravelling tensions and thus they unfold
only partially.\\

\section{Stretching of polyprotein in a flow}

The experiments on protein unfolding in a force clamp are usually performed with
polyprotein chains consisting of several repeats of a given protein.
For example in the
studies of Fernandez group \cite{FernandezLi,Schlierf}, polyubiquitin chains
of 2-9 linked ubiquitin domains were investigated.  When a constant
force is applied
to the terminus of such a system, the domains unfold in a staircase-like manner,
with each step corresponding to the unwinding of a single domain.
Serial unwinding
of polyubiquitin is also observed in molecular dynamics simulations \cite{cs1}.
It is observed that selection of the domain to be unravelled the first
is fluctuations-driven and thus random in nature.\\

In a uniform flow, the situation is different, as it is seen in Figure 10
which presents the unfolding
pathways
for two-ubiquitin for various flow rates. First, because of the
nonuniform tension along the chain,
the unwinding always begins with the domain closest to the the anchoring
point. Also, as it is seen in
Figure 10, if the flow rate is high enough the unfolding is not serial
- one of the ubiquitin domains
unfolds together with a considerable piece of the other domain. For smaller flow rates, one
of the intermediates corresponds to the situation when one of the domains is
fully unfolded while the other is not. However, now it is just one of the many
intermediate states of 2-ubiquitin and not the unique intermediate conformation, as it is
the case for the force-clamp stretching.\\

\section{Non-terminal attachment}

Since unfolding in a uniform flow depends considerably on the choice
of the anchoring terminus, it is worth exploring other possibilities
of anchoring. Figure 11 corresponds to an unfolding trajectory
of integrin chain that is tethered at lysine 148.
Here we monitor the end-to-end lengths, $L_1$ and $L_2$, of
two segments (1-148) and (148-184) respectively, at the
net hydrodynamic force of $\tilde{F}=4$.
We observe that, initially, both segments get streched side-by-side
and at the same rate. However, as
discussed in the Introduction, the tension in a longer segment is
higher which leads to a rapid rupture of the inter-segmental contacts.
From this time on, they evolve independently - the
longer chain unwinds quickly whereas the shorter one snaps back and folds
into a stationary conformation as shown in Figure 11.\\

\section{Elongational flow}

Finally, we observe that the abundance of intermediate states
seen in a uniform flow is not necessarily present for other kinds of flows.
As an illustration, we have carried out
simulations of integrin unfolding in an elongational flow described by:
\begin{equation}
u_x=g (x-x_0), \ \ \ u_y=-\frac{1}{2} g (y-y_0), \ \ \
u_z=-\frac{1}{2} g (z-z_0) \;\;\;.
\end{equation}
Here, $(x_0,y_0,x_0)$ corresponds to a location of the stagnation
point for the flow ({\it cf. }Figure 12)
and we take it to coincide with the C terminus of a protein.
In over 500 unfolding trajectories,
obtained for a broad range of the elongational rate, $g$, we have not observed
even a single stationary intermediate state, whereas there are at least
four intermediate states when flow is uniform and the anchoring is applied
to the C terminus. Figure 13 shows an example of the unfolding pathway in
this case together with the dependence of the mean unfolding time on $g$.\\

The lack of intermediate states in elongational flow, in contrast to
the uniform flow, has been already noticed by Lemak et al. \cite{lemak1,lemak2}.
However, they attributed the absence to the fact that ``in an
elongational flow every
monomer experiences a force that is high enough to delocalize it from
a bonding site''.
In our opinion, the physical mechanism here is, in fact, different.
We note that, in contrast to the case of the uniform flow, the total hydrodynamic force
acting on a protein chain in an
elongational flow depends on the actual total length of the chain. This is
so because the farther from $(x_0,y_0,z_0)$ the pulled terminus is, the
faster flow it experiences. Thus as the protein unravels even just a bit
the total hydrodynamic force also increases correspondingly.
This results in a positive feedback mechanism that leads to a rapid
rupture until the protein is unravelled fully.\\

\section{Refolding after stopping the flow}

If the flow is stopped suddenly, the protein chain folds again. The
analysis of the folding trajectories
shows a considerable number  of misfolded stationary conformations
arising in these
processes (typically about 10-20\% trajectories, depending on the initial
extension  of the protein). Typically, the misfolded conformations lie
relatively close
to the native state, with a RMSD of $1-10 \AA$ but the escape time from the
misfolded conformation is often longer than the time needed
to reach the misfolded state.
A typical trajectory with a misfolding event is presented in Figure 14.
In this trajectory, the protein gets into a conformation
with essentially the same end-to-end length as
in the native state without establishing about 10\% of the native contacts.
The lower panel of
Figure 14 shows the distribution
of the folding times for ubiquitin in which the initial state
has been obtained by a
constant flow unfolding with a total hydrodynamic
force of $\tilde{F}=5$. In this case, almost 20\% of all trajectories
lead to misfolding. The trajectory in the upper panel of
Figure 14 corresponds to a
relatively short-lived misfolded state.
In most situations corresponding to misfolding,
the native conformation is not reached within the duration of the simulation. 
Such trajectories have not been included in the histogram shown in Figure 14.

\section{Summary}

In summary, stretching in a uniform flow provides a promising tool for probing
the conformational landscape of proteins. In contrast to homopolymers,
the steady-state conformations of proteins corresponding to various
flow rates form a discrete set. Unfolding usually involves several kinetic
transitions between subsequent intermediates and has a richer dynamics than that
in the force-clamp case. Moreover, the unfolding pathways depend on the
selection of the point of anchor. Thus, making various selections
provides additional information about the structure of proteins.
Harnessing this information may be facilitated experimentally
by attaching the free end of the protein to a fluorescent quantum dot \cite{Weitz}.
A similar technique  has been succesfully used,
e.g., in tracking of the myosin molecular motor \cite{Warshaw}.\\

M.C. appreciates stimulating discussions with Olle Inganas.
This work was supported by  Ministry of Science in Poland
(Grant No. 2P03B-03225)
and by the European program IP NaPa through Warsaw
University of Technology.

\newpage
\centerline{FIGURE CAPTIONS}

\begin{description}

\item[Fig. 1. ]
Schematic representation of stretching a polymer by fluid flow. The bead
denoted by 1 is free whereas the bead denoted by $N$ is attached to a
spring. The other
end of the spring is fixed. A uniform flow is directed from the right
to the left.
The tension along the chain increases linearly from the free end towards the
anchored end.

\item[Fig. 2. ]
The top-left panel shows the dependence of the fractional extension of
the helix in the steady state on
the total
hydrodynamic force for synthetic helices with $N=48$ (asterisks) and
$N=24$ residues (squares). The snapshots at the bottom show examples of the
stationary states for forces indicated. The top-right panel plots the
logarithm of
the median unfolding time against the total hydrodynamic force for the
$N$=24 helix.
For the purpose of making this figure, we consider
a helix to be unfolded when its total length exceeds 90\% of the
maximum extension length
of $(N-1) \times 3.8 \AA$.

\item[Fig. 3. ]
Examples of the time evolution of the end-to-end distance in unfolding of
integrin in a flow for the total hydrodynamic forces as indicated.
The C-terminal is fixed and the conformations corresponding to the
plateau regimes of the unfolding pathways are shown on the right.

\item[Fig. 4. ]
Similar to Figure 3 but for the situation in which the N terminus is fixed.

\item[Fig. 5. ]
The  scenarios of unfolding of integrin in a uniform flow.
The top and bottom panels corresponds to anchoring of the C and N termini
respectively. The values of the total stretching forces are indicated.

\item[Fig. 6. ]
Similar to Figure 3 but for unfolding induced by applying a constant
force, as indicated, in a force clamp.

\item[Fig. 7. ]
Similar to Figure 3 but for ubiquitin.

\item[Fig. 8. ]
Similar to Figure 4 but for ubiquitin.

\item[Fig. 9. ]
The dependence of the logarithm of the median unfolding time on the force.
The top panel is for ubiquitin and the bottom panel for integrin.
The solid data points and solid lines (the latter are guides to the eye)
correspond to unfolding in a flow. The choice of the anchored
terminus is indicated next to the lines.
The open symbols and dotted lines correspond to stretching in a force clamp.

\item[Fig. 10.]
Unfolding pathways of two-ubiquitin for various total forces as indicated.
An unfolding trajectory for
two-ubiquitin in a force-clamp (the dashed line) is given for comparison.
The full extension in force clamp is longer than in the flow
induced case since in the latter case the segments which are most distant from the
anchor experience too small force to unravel.
The inset shows the scenarios of unravelling events. The contacts
in the domain that is closer to the anchor are marked by black
squares and those in the more distant domain by open squares.
The snapshots on the right correspond to the stationary states
at those values of $L$ (approximately) at which the snapshots are plotted.
The values of dimensionless forces used in the simulations ($\tilde{F}$)  are indicated.

\item[Fig. 11.]
The unfolding pathway of integrin anchored at $Lys 148$ in a uniform
flow. The end-to-end
length of the segments (1-148) and (148-184) is plotted as a function
of time, with the corresponding
protein conformations shown.
The inset shows a protein conformation just before the contacts between the segments are
broken.

\item[Fig. 12. ]
A schemtatic view of elongational flow with a stagnation point $(x_0,y_0)$ in the center of
the graph

\item[Fig. 13. ]
Integrin: the unfolding pathway in an elongational flow and the median
unfolding time as a function of
elongational rate $\tilde{g}$ defined as $\tilde{g}=N \gamma g L_m$,
where $L_m$ is the maximum extension
length of a protein,  $L_m=(N-1) \times 3.8 \AA$. Note that below $g \approx 1.6$
the unfolding process is essentially arrested.

\item[Fig. 14. ]
The top panel shows a typical folding trajectory with a misfold for integrin.
The initial state was obtained by a constant flow unfolding with a
total hydrodynamic
force of $\tilde{F}=5$. The time is measured from an instant at which
the force is
suddenly reduced to zero (i.e. the flow is stopped). Throughout the process, the
C terminus is held anchored. The graph shows both the end-to-end
distance ($L$) and the
fraction of native contacts ($Q$).
Note that the final escape from the misfolded state is only seen in
the $Q(t)$ graph.
The bottom panel shows the distribution of folding times in this case.
The fit is to a log-normal distribution
$\frac{1}{\sqrt{2\pi}\sigma (t-t_0)}
\exp{(-\frac{ln^2(\frac{t-t_0}{m})}{2\sigma ^2})}$.
with $t_0 =7675 \tau$, $\sigma = 0.4$, and $m = 3065 \tau$.
The histogram is based on 300 trajectories all starting from the same stretched
conformation. Trajectories with long-lasting misfolded
conformations (corresponding to the unfolding times longer than $18000
\tau$) have not been taken into account
in the histogram. They make up about 20\% of all trajectories.
\end{description}

\clearpage

\begin{figure}
\epsfxsize=4.6in
\centerline{\epsffile{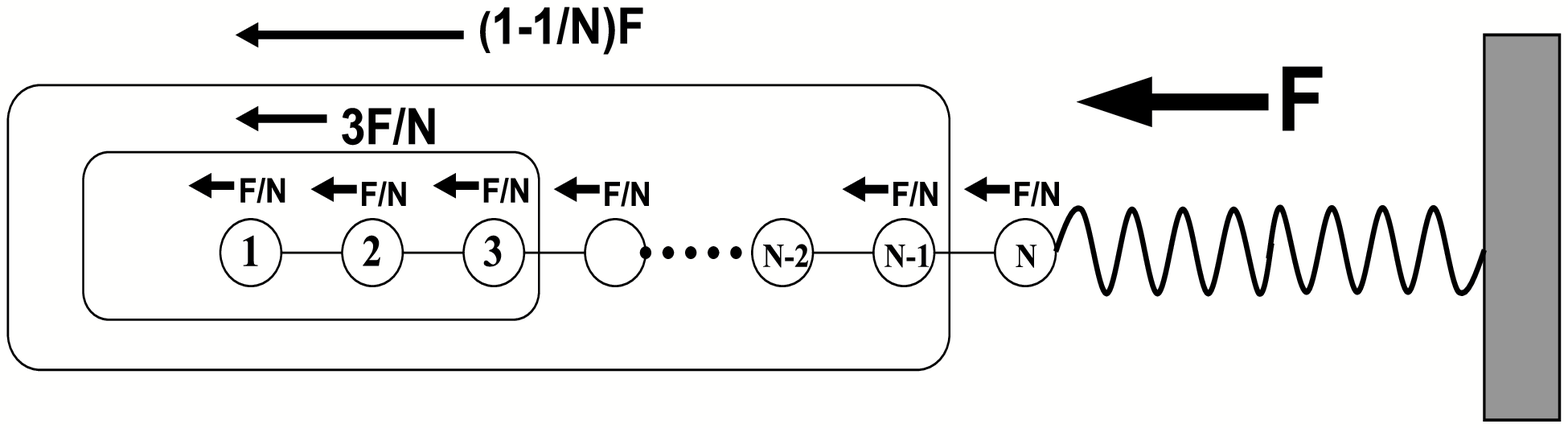}}
\caption{ }
\end{figure}

\vspace*{-4cm}

\clearpage

\newpage

\begin{figure}
\epsfxsize=4.6in
\centerline{\epsffile{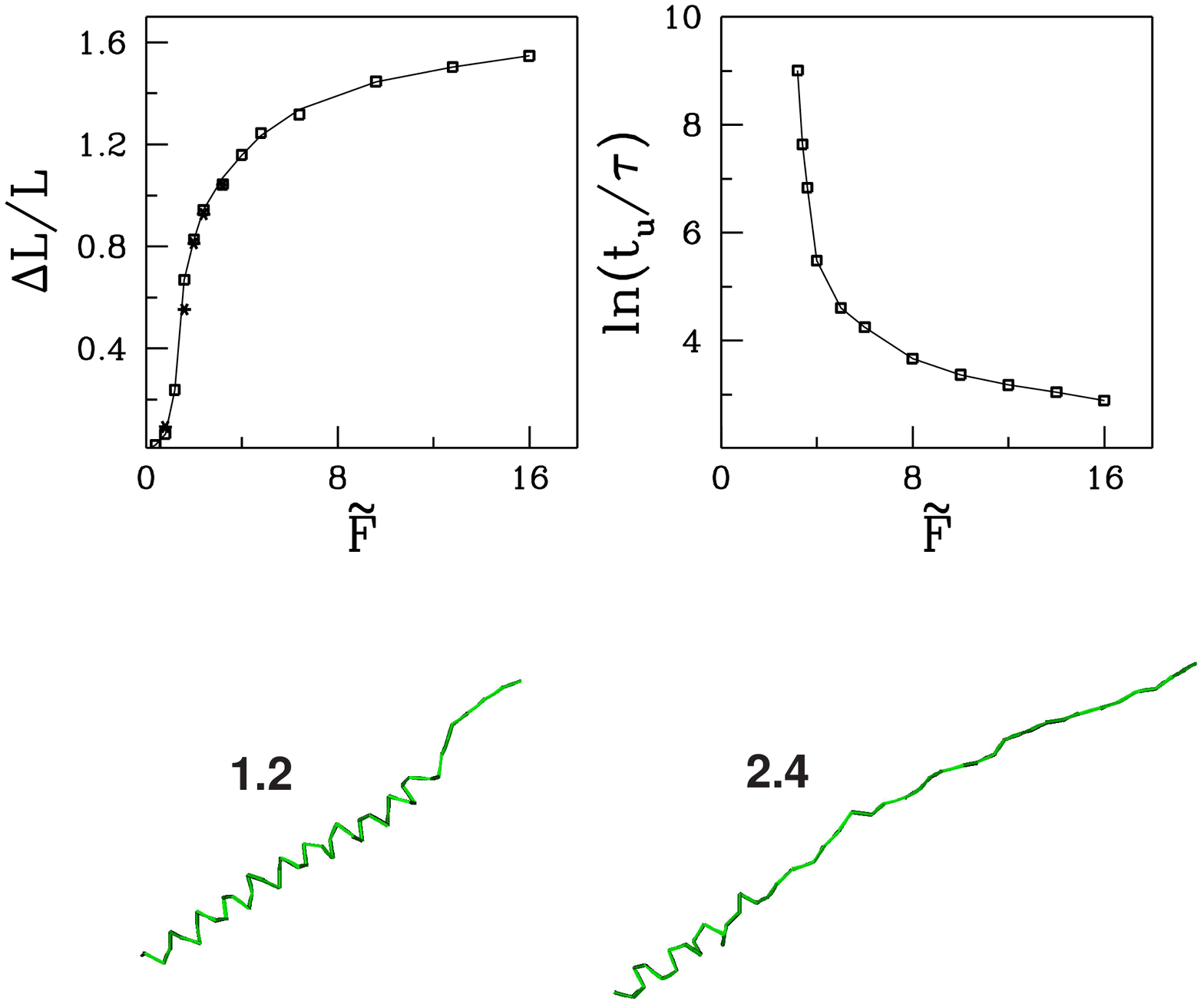}}
\vspace*{-2.5cm}
\caption{ }
\end{figure}

\clearpage

\newpage

\begin{figure}
\epsfxsize=3.8in
\centerline{\epsffile{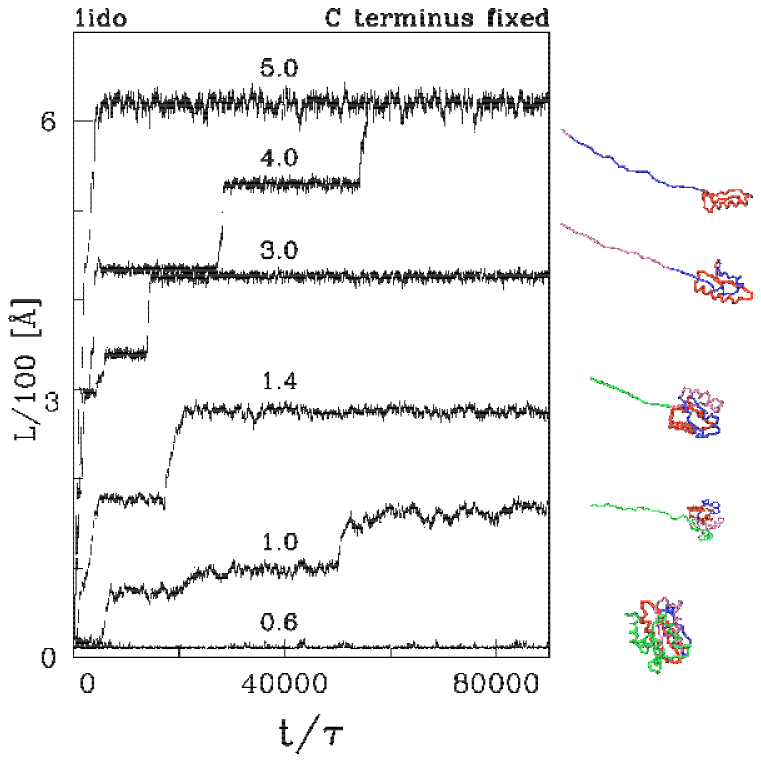}}
\caption{ }
\end{figure}

\vspace*{-4.5cm}

\clearpage

\newpage

\begin{figure}
\epsfxsize=3.8in
\centerline{\epsffile{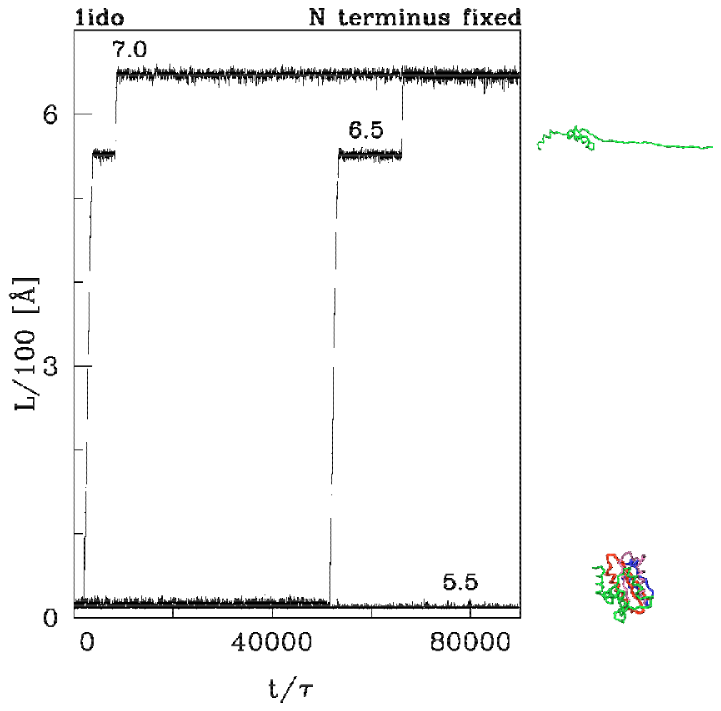}}
\caption{ }
\end{figure}

\clearpage

\newpage

\begin{figure}
\epsfxsize=3.8in
\centerline{\epsffile{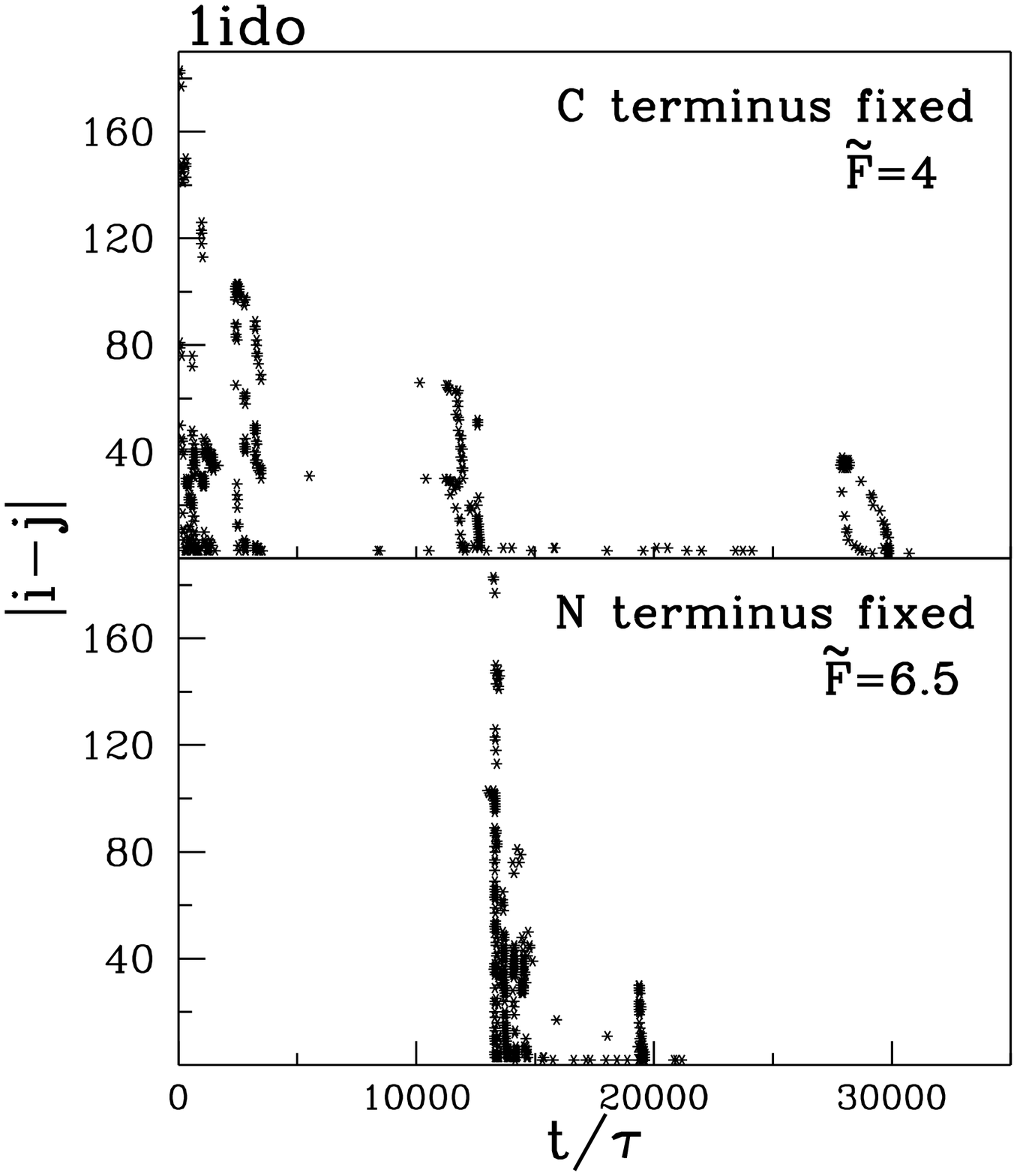}}
\caption{ }
\end{figure}
\vspace{2.5cm}
\clearpage

\newpage

\begin{figure}
\epsfxsize=3.8in
\centerline{\epsffile{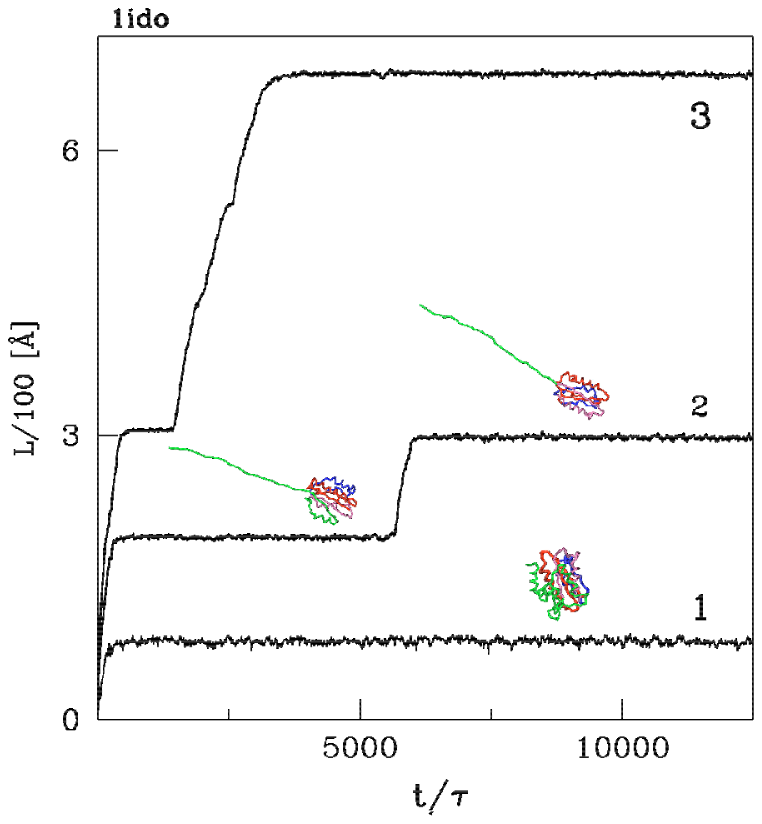}}
\caption{ }
\end{figure}

\clearpage

\newpage

\begin{figure}
\epsfxsize=3.8in
\centerline{\epsffile{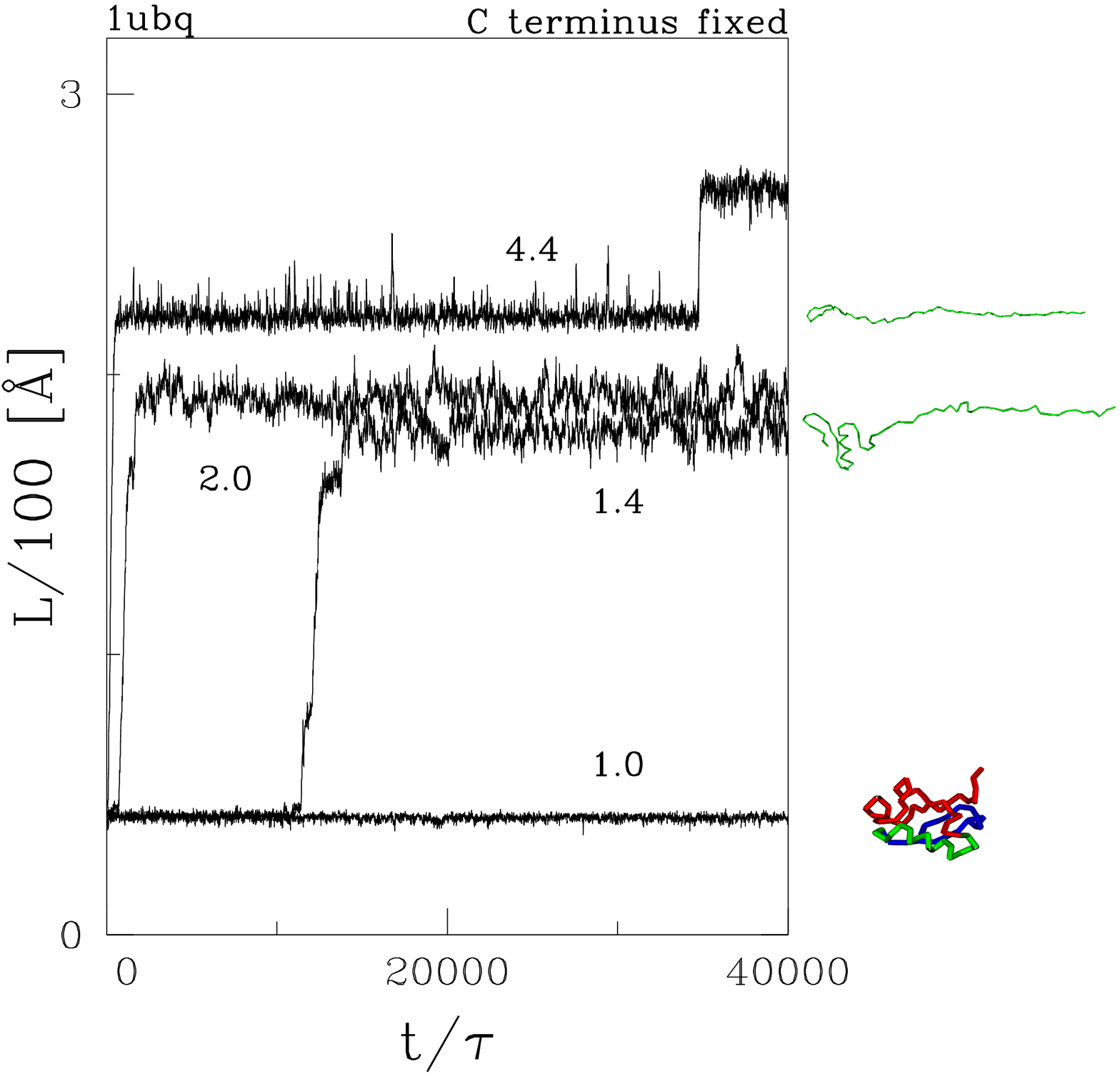}}
\caption{ }
\end{figure}

\clearpage

\newpage

\begin{figure}
\epsfxsize=3.8in
\centerline{\epsffile{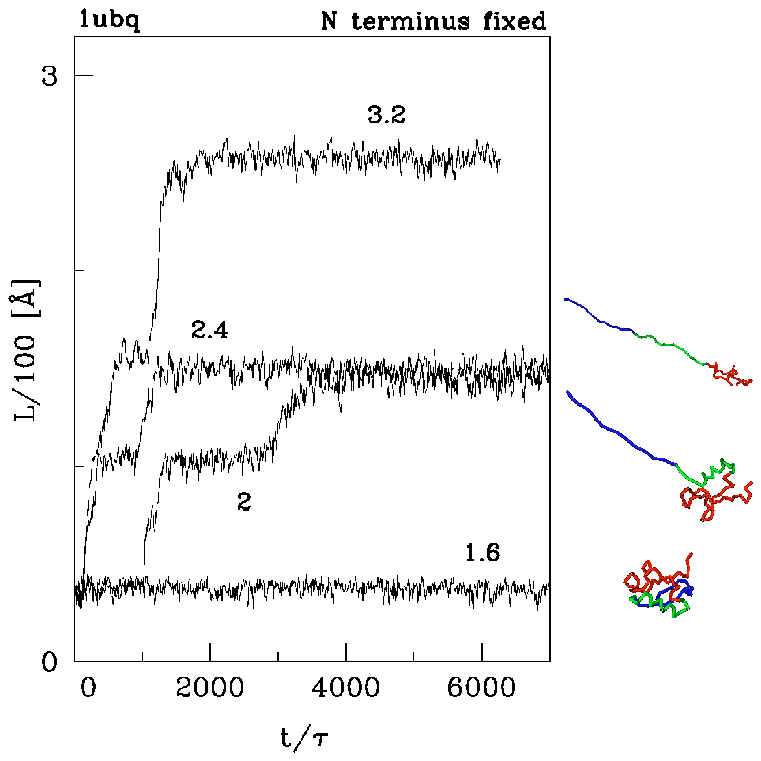}}
\caption{ }
\end{figure}

\clearpage

\newpage

\begin{figure}
\epsfxsize=3.8in
\centerline{\epsffile{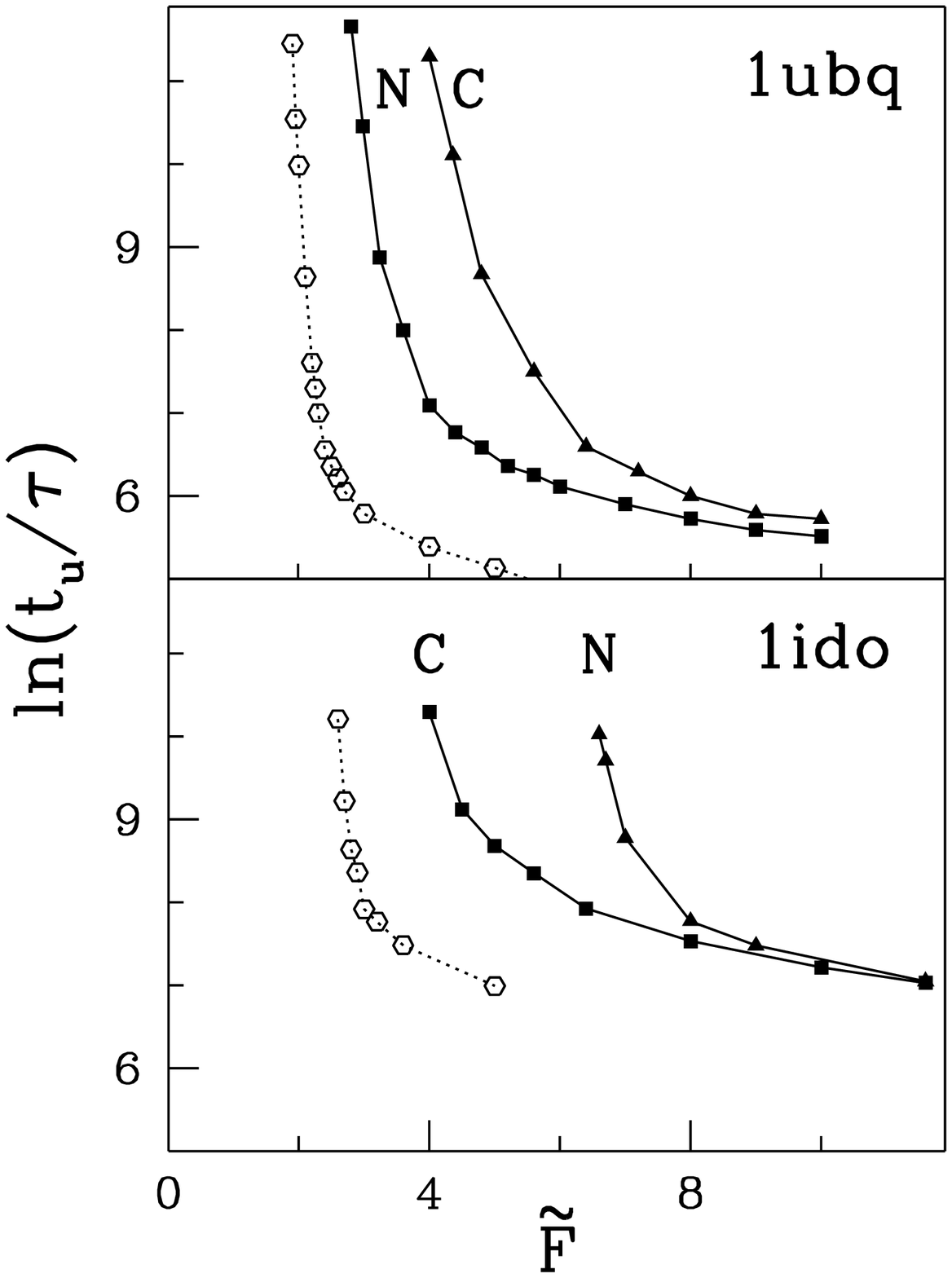}}
\vspace*{-0.2cm}
\caption{ }
\end{figure}

\clearpage

\newpage

\begin{figure}
\epsfxsize=3.8in
\centerline{\epsffile{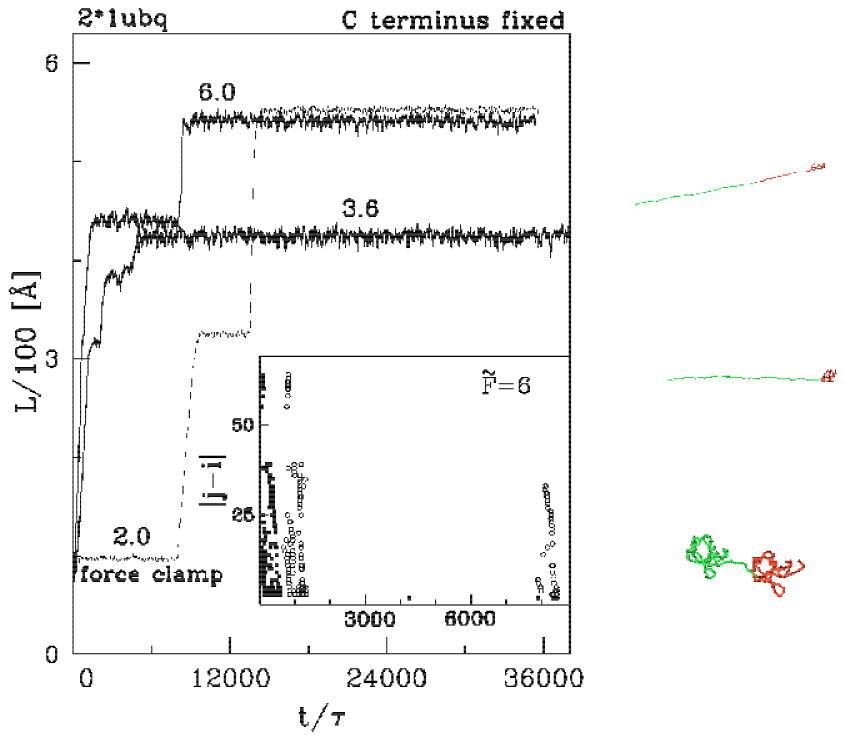}}
\caption{ }
\end{figure}

\clearpage


\begin{figure}
\epsfxsize=3.8in
\centerline{\epsffile{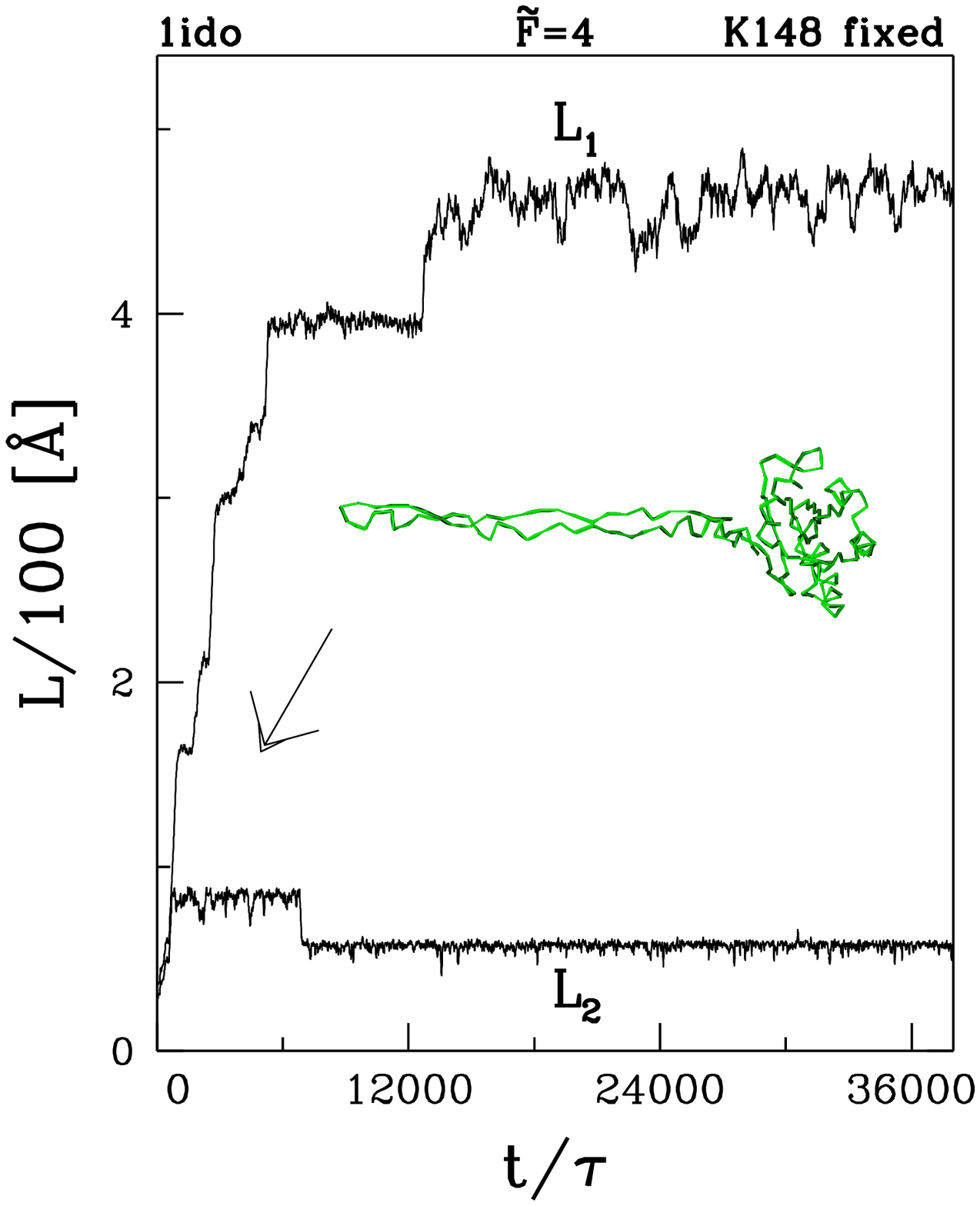}}
\vspace*{-2cm}
\caption{ }
\end{figure}

\clearpage

\newpage

\begin{figure}
\epsfxsize=2.8in
\centerline{\epsffile{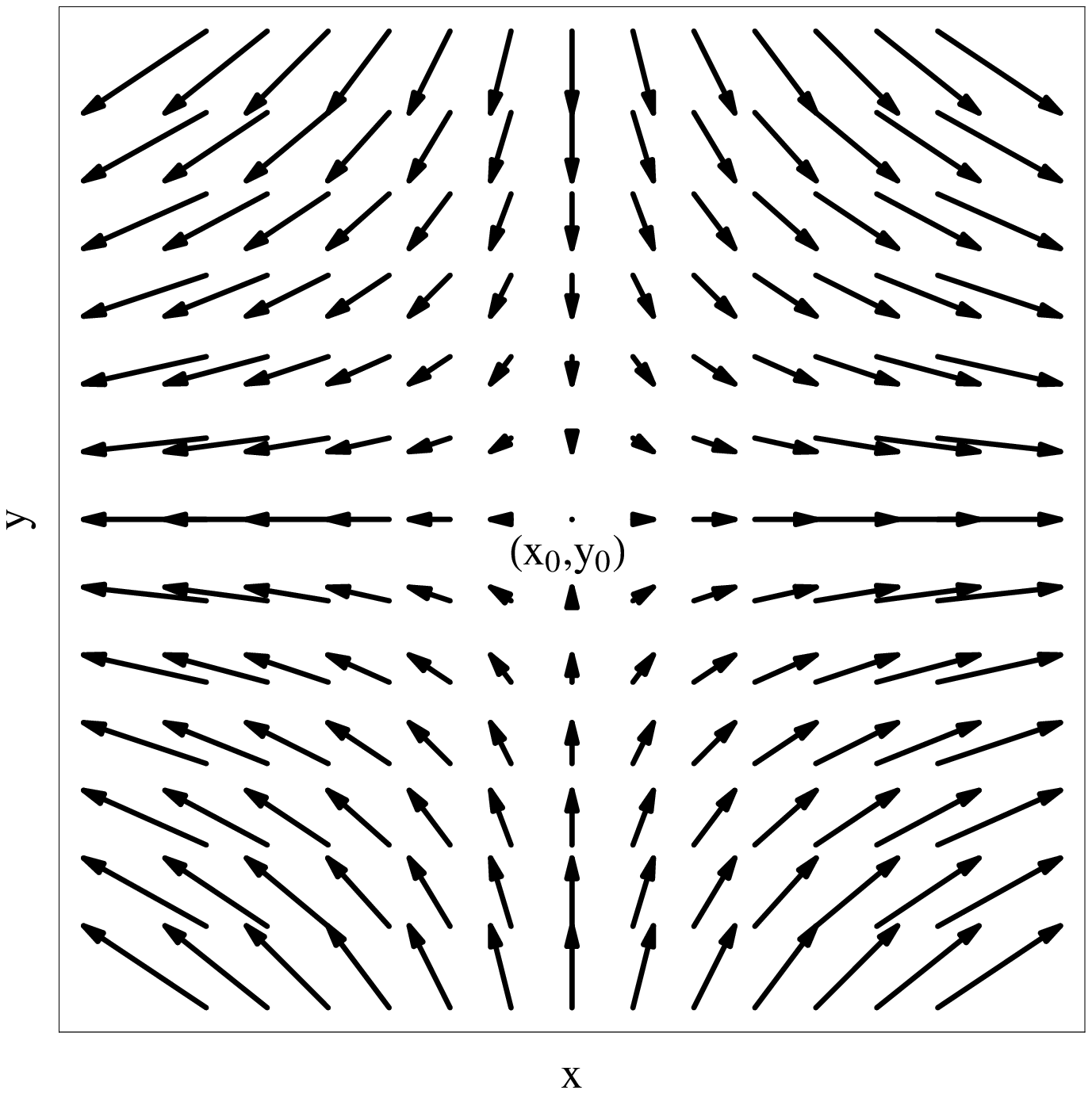}}
\caption{ }
\end{figure}

\clearpage
\newpage

\begin{figure}
\epsfxsize=3.8in
\centerline{\epsffile{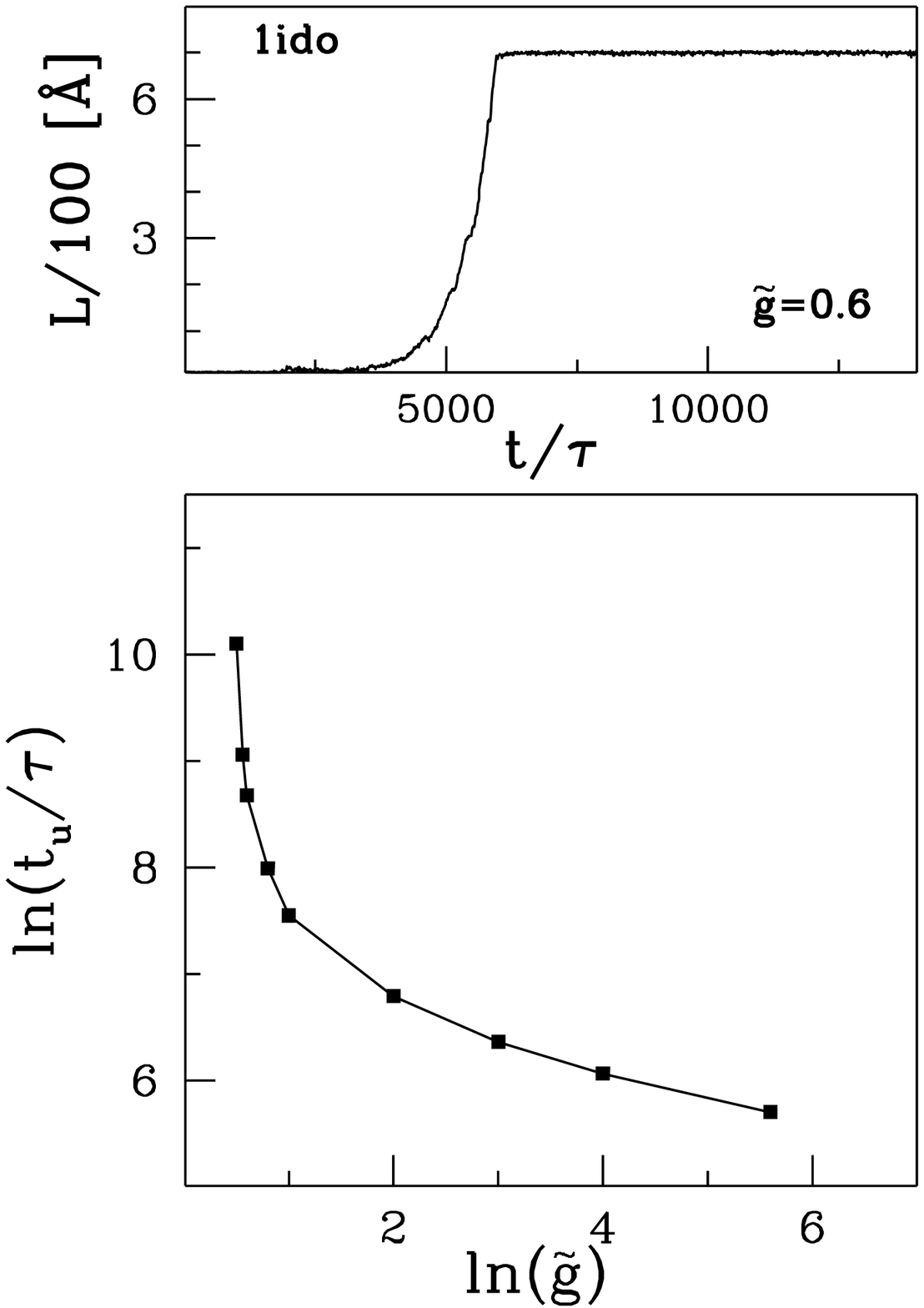}}
\caption{ }
\end{figure}

\clearpage

\newpage

\begin{figure}
\epsfxsize=3.9in
\centerline{\epsffile{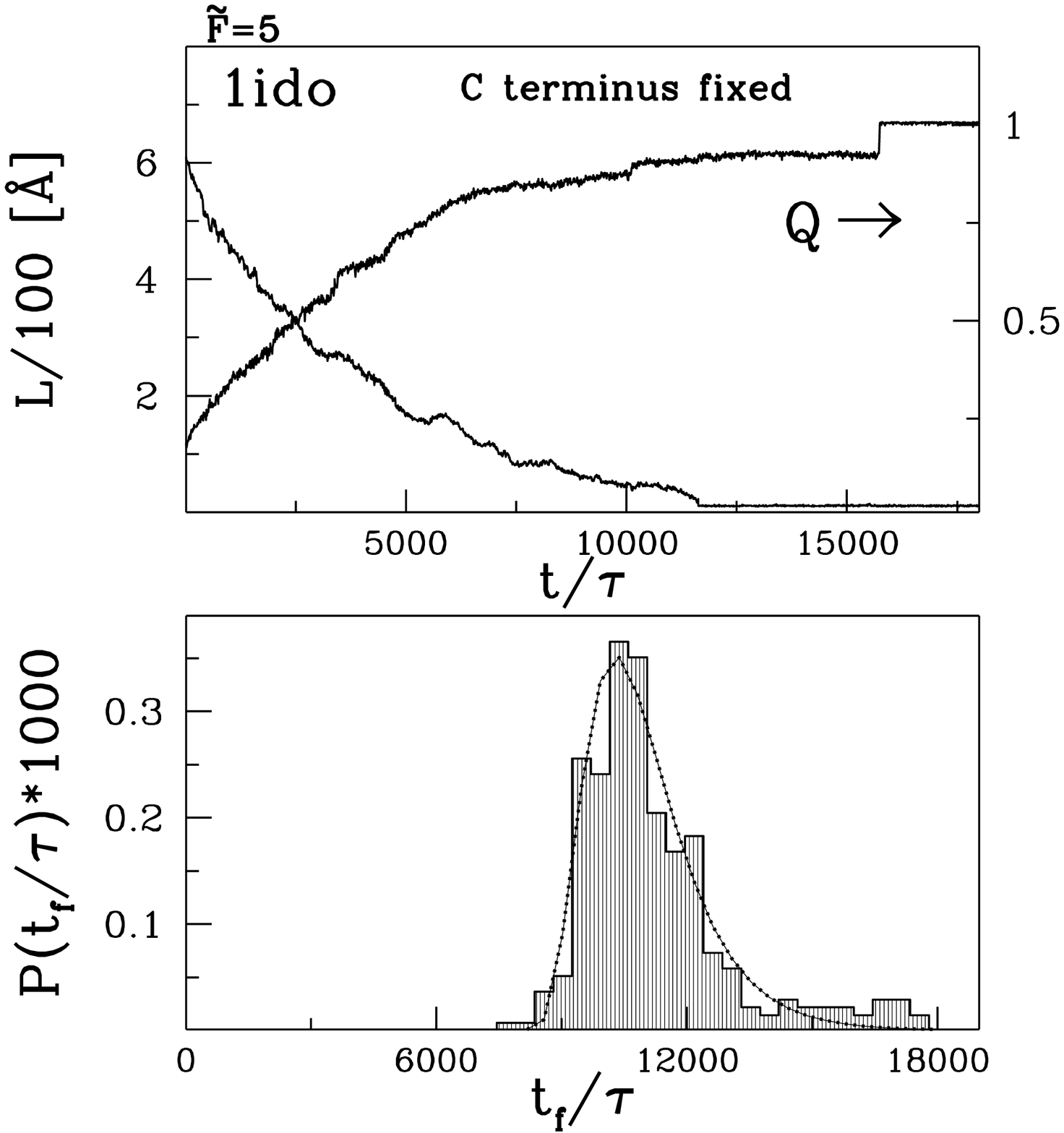}}
\caption{ }
\end{figure}

\clearpage

\end{document}